\documentclass{aa}

\usepackage{txfonts}
\usepackage{graphicx}
\usepackage{natbib}
\bibpunct{(}{)}{;}{a}{}{,}

\begin{document}

\title{Designing Future Dark Energy Space Missions: I. Building
Realistic Galaxy Spectro-Photometric Catalogs and their first
applications}

\author{ S. Jouvel \inst{1} \and J-P. Kneib \inst{1} \and O. Ilbert
\inst{3,1} \and G. Bernstein \inst{2} \and S. Arnouts \inst{1,4}
\and T. Dahlen \inst{5} \and A. Ealet \inst{9,1} \and B. Milliard
\inst{1} \and H. Aussel \inst{7}\and P. Capak \inst{6} \and
A. Koekemoer \inst{5} \and V. Le Brun \inst{1} \and H. McCracken
\inst{8} \and M. Salvato \inst{6} \and N. Scoville
\inst{6}}\institute{Laboratoire d'Astrophysique de Marseille,
CNRS-Universit\'e d'Aix-Marseille, 38 rue Frederic Joliot-Curie;
13388 Marseille cedex 13, France \and University of Pennsylvania,4N1
David Rittenhouse Lab 209 S 33rd St Philadelphia, PA 19104, USA \and
Institute of Astronomy, 2680 Woodlawn Drive Honolulu, HI 96822-1897,
USA \and CFHT, 65-1238 Mamalahoa Hwy Kamuela, Hawaii 96743 USA \and
Space Telescope Science Institute, 3700 San Martin Drive Baltimore,
MD 21218 USA \and California Institute of Technology 1200 East
California Blvd. Pasadena CA 91125, USA \and Service
d'Astrophysique, CEA-Saclay, 91191 Gif-sur-Yvette, France \and
Institut d'Astrophysique de Paris 98bis, bd Arago F75014 Paris,
France \and Centre de Physique des Particules de Marseille, 163,
avenue de Luminy, Case 902, 13288 Marseille cedex 09, France}

\date{today , accepted later}

\authorrunning{St\'ephanie Jouvel et al. } \titlerunning{Mock Galaxy
Catalogs for DE} \offprints{Stephanie Jouvel,
\email{stephanie.jouvel@oamp.fr}}

\abstract{
Future dark energy space missions such as JDEM and EUCLID are being
designed to survey the galaxy population to trace the geometry of
the universe and the growth of structure, which both depend on the
cosmological model. To reach the goal of high precision cosmology
they need to evaluate the capabilities of different instrument
designs based on realistic mock catalogs of the galaxy distribution.
}{
The aim of this paper is to construct realistic and flexible mock
catalogs based on our knowledge of galaxy populations from current
deep surveys. We explore two categories of mock catalogs : (i) based
on luminosity functions that we fit to observations (GOODS,
UDF,COSMOS,VVDS) (ii) based on the observed COSMOS galaxy
distribution. }{
The COSMOS mock catalog benefits from all the properties of the
data-rich COSMOS survey and the highly accurate photometric redshift
distribution based on 30-band photometry. Nevertheless this catalog
is limited by the depth of the COSMOS survey. Thus, we also evaluate
a mock galaxy catalog generated from luminosity functions using the
\emph {Le Phare} software. For these two catalogs, we have produced
simulated number counts in several bands, color diagrams and redshift
distributions for validation against real observational data.}{
Using these mock catalogs we derive some basic requirements to help
design future Dark Energy missions in terms of the number of galaxies
available for the weak-lensing analysis as a function of the PSF size
and depth of the survey. We also compute the spectroscopic success
rate for future spectroscopic redshift surveys \emph {(i)} aiming at
measuring BAO in the case of the wide field spectroscopic redshift
survey, and \emph {(ii)} for the photometric redshift calibration
survey which is required to achieve weak lensing tomography with great
accuracy. In particular, we demonstrate that for the photometric
redshift calibration, using only NIR (1-1.7$\mu$m) spectroscopy, we
cannot achieve a complete spectroscopic survey down to the limit of
the photometric survey ($I<25.5$).  Extending the wavelength coverage
of the spectroscopic survey to cover 0.6-1.7$\mu$m will then improve
the fraction of very secure spectroscopic redshifts to nearly 80\% of
the galaxies, making possible a very accurate photometric redshift
calibration.  }{
We have produced two realistic mock galaxy catalogs that can be used
in determining the best survey strategy for future dark-energy
missions in terms of photometric redshift accuracy and spectroscopic
redshift surveys yield. {\it These catalogues are publicly accessible
  at http://lamwws.oamp.fr/cosmowiki/RealisticSpectroPhotCat, or by
  request to the first author of this paper.}  }

\date{\fbox{\sc Draft Version: \today}}
\maketitle

\section{Introduction}
\label{sec:intro}
The prospect for high-precision cosmological inferences from large
galaxy surveys has prompted the initiation of several projects with
the goal of surveying thousands of square degrees of sky in multiple
filters with or without spectroscopy. The ground based projects
[e.g. current KIDS, DES, and future Pan-STARRS, LSST in imaging and
  current SDSS-III/BOSS and future WFMOS BigBOSS in spectroscopy], and
the space based missions [JDEM, and EUCLID, or their former and future
  concepts] all propose to conduct wide field galaxy survey (in
imaging with or without spectroscopy) in order to exploit the power of
weak gravitational lensing [WL] and galaxy clustering (using baryonic
acoustic oscillations [BAO] with or without redshift distortion
measurements [RD]) to elucidate the cause of the acceleration of the
Hubble expansion (\citet{Riess98}, \citet{Perlmutter99},
\citet{Astier06}, \citet{Kilbinger09}). Proper design and forecasting
of the performance of these experiments require an accurate estimate
of the ``yield'' of galaxies, such as number counts, redshift sizes
and color distributions, from a chosen survey configuration. It is
typically straightforward to estimate the expected resolution and
noise properties of the telescope and instrument, but more difficult
to quantify the number and properties of the ``useful'' galaxies
available on the sky. A good forecast requires that we estimate the
density of galaxies on the sky over the joint distribution of:\\ (i)
redshift; (ii) angular size, expressed as half-light radius $r_{1/2}$;
(iii) apparent magnitudes and colors in any chosen instrument
passbands; (iv) emission-line strengths.\\ The first three properties
are essential to knowing whether a given galaxy will be detected at
sufficient signal-to-noise ($S/N$) and resolution to determine its
shape and its photometric redshift. The last property, the
emission-line strength, is essential to estimating the depth and
completeness that any spectroscopic redshift survey will
achieve. Spectroscopic redshift accuracy is needed to conduct the best
possible BAO/RD measurements, and to calibrate photometric redshift
(``photo-z'') estimators that are essential for WL tomographic
measurement (e.g. \citet{Massey07}).

In this paper we present two simulated catalogs of galaxy properties
based on current deep surveys that we will use in forthcoming papers
(Jouvel et al. 2009 in prep.) to forecast the performance of WL Dark
Energy space based missions. Unfortunately we cannot simply use a
catalog from some completed survey, since (1) no single survey of
useful depth has simultaneously observed all of the listed properties
for its target galaxies, and (2) the proposed surveys will exceed the
depth, field of view, resolution, with or without wavelength coverage
of most existing observed galaxy catalogs. For example, the space based
spectroscopic survey will likely mainly be conducted in the
near-infrared, surpassing any current infrared spectroscopic redshift
survey. The mission concepts of SNAP, Destiny, and EUCLID propose
near-infrared (NIR) imaging over very wide areas, but only a few square
arcminutes of HST/NICMOS imaging data with the UDF \citep{Coe06} and
close to one degree down to $K\sim23$ for ground based data with the
UKIDSS survey on the UKIRT telescope \citep{Lawrence07} are available
to date at these magnitude levels, insufficient to serve as a robust
source model. Any relevant simulated catalogs must therefore include
some degree of extrapolation or modelling of the source population.

In this paper, we explain how we have constructed 2 simulated galaxy
catalog based on deep survey data. Section \ref{sec:mock} explain the
methodology used. In section \ref{sec:validate} we validate these
simulated catalogs by comparing their predicted magnitude, color,
redshift, size, and emission-line strength distributions to real
survey data taking into account the survey selection functions.

Although this work was initiated in the context of the SNAP
collaboration (see our first results in \citet{Dahlen08}), it can
easily be adapted to any instrument concept for a proper evaluation of
its merit. When necessary we assume a $\Lambda CDM $ universe:
$(\Omega_{M},\Omega_{\Lambda})=(0.3,0.7)$ and $H_0=70$ km/s/Mpc. All
magnitudes used in this paper are on the AB system.

\section{Realistic mock Galaxy Catalog}
\label{sec:mock}
For both of our simulated catalogs, each galaxy is assigned a spectral
energy distribution (SED), which can be integrated over any
instrumental passband to forecast an apparent magnitude. Our first
fully simulated catalog has been generated by using ``Le Phare''
simulation tool \footnote{Arnouts
http://www.oamp.fr/people/arnouts/LE\_PHARE.html} (Arnouts and
Ilbert 2009 in prep) with an analytic luminosity function based on
the GOODS survey from the work of \citet{Dahlen05}. We will refer to
our derived simulated catalog as the {\em GOODS Luminosity Function
Catalog (GLFC)}.
\subsection{GLFC - GOODS Luminosity Function based Catalog}
\subsubsection{SED Library}
\label{subsec:library}
\begin{figure}
\resizebox{\hsize}{!}{\includegraphics{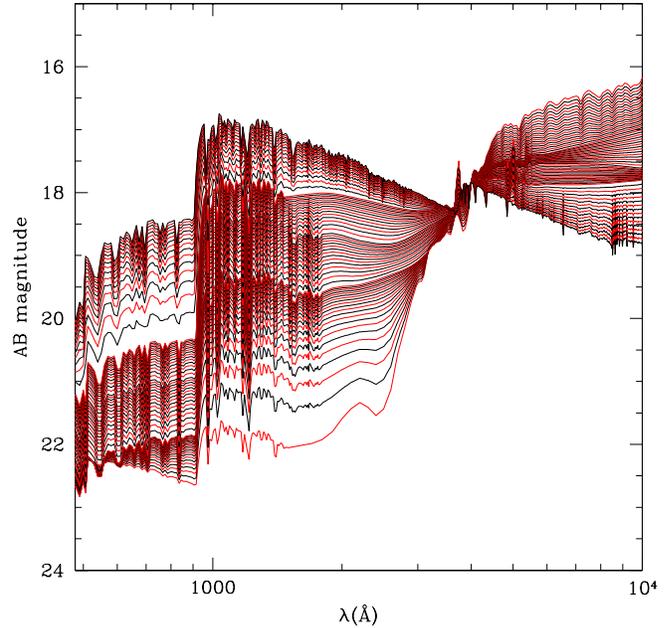}}
\caption{Extended CWW library of SED templates. This represents the
full range of SED template linearly interpolated between the 4 CWW
templates and one star-forming template in AB magnitude (arbitrarily
flux scaled at 4000\AA).}
\label{fig:library}
\end{figure}

\begin{figure}
\resizebox{\hsize}{!}{\includegraphics{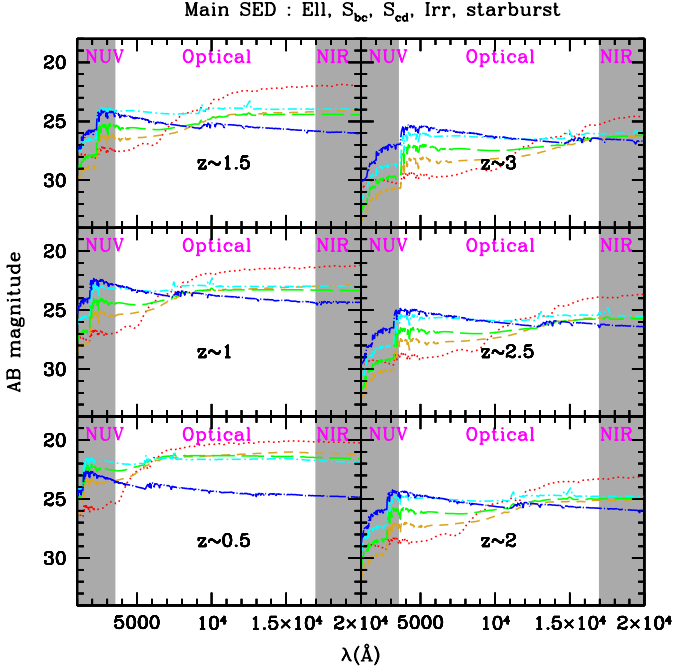}}
\caption{Magnitudes of the 5 main templates composed by the 4 CWW :
  $Ell$ (red dotted), $S_{bc}$ (gold dashed), $S_{cd}$ (green
  long-dashed), $Irr$ (cyan dot-dashed) and one star-forming (blue
  long dot-dashed) shown at some given redshift of
  0.5,1,1.5,2,2.5,3. The templates are $M*$ galaxies as given in
  Table~\ref{LF}.}
\label{fig:library2}
\end{figure}

The GLFC relies on two ingredients :
\begin{itemize}
\item A set of SEDs spanning the entire range from Elliptical to
starbusting galaxy,
\item A redshift evolving luminosity function (LF) per type.
\end{itemize}
Those ingredients are the basic requirements to generate a catalog
that can predict global distributions such as magnitude, redshift and
color counts. We use the Coleman Extended library (CE) for our
templates of galaxies. This list is based on the four observed spectra
of \citet{Coleman80} corresponding to an Elliptical, Sbc, Scd and
Irregular, which have been extrapolated in the UV and IR wavelengths
domain by using synthetic spectra from the GISSEL library (Charlot and
Bruzual,1996). To reproduce observed colors bluer than the CE
templates, we add one spectrum of star-forming galaxies computed with
the GISSEL model for solar metallicity, Salpeter IMF, constant star
formation rate and 0.05 Gyr age. Following the approach adopted by
\citet{Sawicki96}, we have linearly interpolated the original 5 SEDs
to provide a finer grid of spectral-type coverage with a total library
of 66 templates shown in {\bf Figure~\ref{fig:library}}. In {\bf
  Figure~\ref{fig:library2}}, we draw magnitudes of the 5 main
templates in a magnitude-wavelength plane by bins of redshifts in
order to show colors of these templates that may help in deciding
which sensitivities are needed to detect galaxies inside different
passbands.
\begin{table}[!ht]
\caption{Restframe colors of SED templates.}
\begin{tabular}{cccccccc} \hline\hline
LF & E(B-V) & $types$ & $B-V (restframe)$ \\
1 & 0 & 1 $\to$ 17 & 1.069 $\to$ 0.701 \\
2 & 0 & 17 $\to$ 55 & 0.701 $\to$ 0.2167\\
3 & 0 & 55 $\to$ 66 & 0.2167 $\to$ -0.0301\\
\textbf{3} & 0.1 & 56 $\to$ 66 & 0.3252 $\to$ 0.1016\\
\textbf{3} & 0.2 & 56 $\to$ 66 & 0.4592 $\to$ 0.2345\\
\textbf{3} & 0.3 & 56 $\to$ 66 & 0.5945 $\to$ 0.3686\\
\hline\hline
Main templates & E(B-V) & nbr & $B-V (restframe)$ \\
$Ell$ & 0 & 1 & 1.069 \\
$S_{bc}$ & 0 & 21 & 0.6262 \\
$S_{cd}$ & 0 & 36 & 0.5077 \\
$Irr$ & 0 & 51 & 0.3185 \\
$star-forming$ & 0 & 66 & -0.0301\\
\hline\hline
\label{BVrest}
\end{tabular}
\end{table}
\subsubsection{Extinction}
We have diversified our templates by adding extinction computed from
the Calzetti extinction law \citep{Calzetti00} with the excess
redenning values $E(B-V)$ of 0,0.1,0.2,0.3 mag and using the
extinction formula:
\begin{eqnarray}
flux_{attenuated}(\lambda)=flux_{intrinsic}(\lambda).10^{-0.4*k(\lambda).E(B-V)}
\end{eqnarray}
The $k(\lambda)$ values follow the Calzetti law. Each $E(B-V)$ value
will be applied to the irregular and
star-forming templates, increasing the number of templates for those
types. Finally we also include the Lyman absorption by the Intergalactic
Medium as a function of redshift, following \citet{Madau95}.

\subsubsection{GLFC Luminosity Function}
\label{subsec:GOODSLF}
The GLFC is generated by assuming a luminosity function (LF) for
galaxies of 3 different types, then drawing galaxies from these LFs to
populate our simulated sky area. We have selected these 3 types
following the B-V rest-frame colors given in \cite{Dahlen05} (see
table~\ref{BVrest}). We start with the LF estimated in rest-frame $B$
band by \citet{Dahlen05} in the 3 spectral types inside 4 redshifts
bins over $0<z<1$. \citet{Dahlen05} derive these LFs by fitting to
photo-z data from the GOODS survey. We have extrapolated these LFs to
$z=6$ to be complete for the galaxies likely to be used in future
surveys. The Dahlen $z=1$ LF for each type is held nearly fixed for
$1<z<6$, with some adjustments made to improve the match to data from
the COSMOS survey (described below).

Table~\ref{LF} gives the LF calculated for 3 galaxy types extrapolated
until redshift $z\simeq 6$ as explained above. To produce the mock
catalog, knowing the luminosity function, Le Phare derives a number of
objects by magnitude and redshift bins $(z,m)$ using a Schechter
function (Schechter 1976) :
\begin{equation}
n(M(z,m))dM=\phi^{*} \left( \frac{M(z,m)}{M^{*}} \right) ^{\alpha} exp\left(-\frac{M(z,m)}{M^{*}}\right)\frac{dM}{M^{*}}
\end{equation}
M is the absolute magnitude which is a function of redshift and
apparent magnitude $(z,m)$, $M^{*}$,$\phi^{*}$ and $\alpha$ are the
parameters of the Schechter function given in the Table \ref{LF}. \\

According to the adopted area, the simulated galaxy counts in each
magnitude-redshift-type bin are drawn from a Poisson distribution of
the expected number based on the LF.

Note that the luminosity function parameters depend on the
cosmological model assumed (as the luminosity function is expressed in
physical units). However the resulting galaxy catalog can be
considered to be independent of the input cosmological parameters
assumed here, as it is built to reproduce the {\em observed} galaxy
distribution and properties.

\begin{table}[!ht]
\caption{Luminosity Function in the B band extrapolated to match COSMOS colors and number counts used to create the GLFC catalog.}
\begin{tabular}{cccccccc} \hline\hline
LF &	$z$ &	$\phi^{*}$ &	$M^{*}$ &	$\alpha$ \\ 
1 &	0.5 &	16.7e-4	&	-21.15 &	-0.71 \\ 
1 &	0.75 &	18e-4	&	-21.13 &	-0.50 \\ 
1 &	1.0 &	9.5e-4	&	-21.72 &	-0.98 \\ 
1 &	2.0 &	8.5e-4	&	-21.72 &	-0.98 \\ 
1 &	3.0 &	7.5e-4	&	-21.72 & 	-0.98 \\ 
1 & 	6.0 & 	6.5e-4	&	-21.72 & 	-0.98 \\ 
2 & 	0.5 &	24.0e-4	&	-21.08 & 	-1.37 \\ 
2 & 	0.75 &	25.0e-4	&	-21.29 & 	-1.17 \\ 
2 &	1.0 & 	25.0e-4	&	-21.2 & 	-1.1 \\ 
2 &	1.5 &	24.4e-4	&	-21.15 & 	-0.05 \\ 
2 &	2.0 &	24.4e-4	&	-21.15 & 	-0.99 \\ 
2 & 	3.0 &	24.4e-4	&	-21.15 & 	-0.99 \\ 
2 &	6.0 &	24.4e-4	&	-21.15 & 	-0.85 \\ 
3 &	0.5 &	48.6e-4	&	-18.84 & 	-1.1 \\ 
3 &	0.75 &	33.1e-4	&	-19.67 &	-1.18 \\ 
3 &	1.0 &	33.1e-4	&	-20.55 &	-1.52 \\ 
3 &	1.5 &	30.0e-4	&	-20.62 &	-1.62 \\ 
3 &	2.0 &	25.0e-4	&	-20.72 &	-1.7 \\ 
3 &	3.0 &	25.0e-4	&	-20.72 &	-1.8 \\ 
3 &	6.0 &	25.0e-4	&	-20.72 &	-1.8 \\ 
\hline 
\hline
\label{LF}
\end{tabular}
\end{table}

Drawing from the luminosity functions gives the galaxy distribution
over the joint magnitude-color-redshift space. We assign a half-light
radius to each galaxy as follows: we first calculate the galaxy's
apparent magnitude in the F814W HST filter. \citet{Leauthaud07}
provides a size-magnitude catalog for the 1.64 deg$^2$ COSMOS survey
executed in this filter. We assign the simulated galaxy a half-light
radius that is drawn at random from all galaxies of the same F814W
apparent magnitude in the observed COSMOS catalog. Thus the
size-magnitude distribution of the simulated catalog will, by
construction, exactly match the COSMOS observations. The procedure
does not reproduce any additional dependence of galaxy size upon type,
color, or redshift using the COSMOS catalog. Any further correlation
between size and other parameters such as color, redshift and galaxy
type could in principle be implemented by following the COSMOS
catalogue. As this is not needed in this paper, we have not
implemented these higher order correlations. However the
size-magnitude relation of the mock catalogue do follow the COSMOS
relation.
 
\subsubsection{Photometric Noise}
\label{subsubsec:noise}
A real survey has noise from the finite photon counts and detector
noise. Estimation of this noise is of course essential for forecasting
survey performance. It may also be important for validation of the
simulation against real data, since the noise can induce biases on
number counts. Le Phare produces both noiseless magnitudes and noisy
magnitudes for each chosen observation bandpass. The noisy magnitudes
are randomly drawn from a Gaussian distribution with mean and standard deviation
$(m,err_m)$. We define a reference couple $(m^*,err^*)$.
At magnitude $m<m^*$, the object noise is dominant and we
assume that the magnitude error follows a power law of the adjustable
form :
\begin{equation}
err_m = 10^{(0.4(p+1)(m-m^*))}
\end{equation}
At magnitude $m>m^*$, the sky noise is dominant and we assume the
error on magnitude follows an exponential law :
\begin{equation}
err_m = err^*/2.72.exp(10^{(q(m-m^*))})
\end{equation}
$(p,q)$ are the slopes for the power laws, both derived from the
survey characteristics.

We will, however, use noiseless magnitudes for the validation tests in
this paper, because the comparison surveys ({\it e.g.} the UDF) have
high $S/N$ in the regimes of comparison, and because the noise-induced
biases are generally small.


\subsection{CMC - The COSMOS Mock Catalog}
\label{subsec:CMC}
The second simulated galaxy catalog is built directly from the
observed COSMOS catalog of Capak et al. (2009) in prep. and
\citet{Ilbert09}. We will refer to this simulation as the COSMOS mock
Catalog (CMC).

\subsubsection{The COSMOS catalog}

The COSMOS photometric-redshift catalog \citep{Ilbert09} was computed
with 30 bands over $\sim$2-deg$^2$ taken from GALEX for UV bands,
Subaru for the optical (U to z), and CFHT, UKIRT and Spitzer for the
NIR bands. However, we restrict our mock catalog to the central square
area of 1.38 deg$^2$ which is fully covered by HST/ACS imaging. The
COSMOS-ACS catalog gives 592000 galaxies for an area of 1.38deg$^2$
. This is roughly a density of 120 galaxies per arcmin$^2$ down to
$i^+ <26.5$. In the COSMOS photometric-redshift catalog, $10\%$ of
this surface corresponds to areas masked because of bright stars that
prevent quality multiband photometry in the extended bright star
halos. The effective area is thus in fact 1.24 deg$^2$ of unmasked
region with a total number of 538000 simulated galaxies out to $i^+
<26.5$ leading again to roughly a density of 120 gal/arcmin$^2$. Point
sources such as stars and X ray sources (mostly dominated by an AGN)
were also removed from the mock catalog.

The photo-z accuracy is based on a comparison to spectroscopic surveys
like the zCOSMOS (bright survey down to $i^+_{med}<22.5$ with 4148
galaxies and faint down to $i^+_{med}<24$ with 148 galaxies) and also
the MIPS infrared selected sample (bright and faint with 317 galaxies,
Figure 7 and 8 of \citet{Ilbert09}).  The COSMOS photo-z accuracy is
$\sigma_{\Delta z/(1+z_{\rm s})}=0.007$ at $i^+<22.5$ with a
catastrophic rate below $1\%$ (see Figure 6 of \citet{Ilbert09}). At
fainter magnitudes and $z<1.25$, the estimated accuracy is
$\sigma_{\Delta z}=0.02, 0.04, 0.07$ at $i^+ \sim 24$, $i^+ \sim 25$,
$i^+ \sim 25.5$, respectively. The accuracy is degraded at
$i^+>25.5$. The deep NIR and IRAC coverage enables the photo-z to be
extended to $z\sim2$ albeit with a lower accuracy ($\sigma_{\Delta
  z/(1+z_{\rm s})}=0.06$ at $i^+_{\rm AB}\sim 24$) (see \citet{Ilbert09}
for more details).

Despite the lower photo-z accuracy at $i^+>25.5$ and $z>1.25$ and the
possible bias due to the faint AGN contribution, we use the full
COSMOS catalog. Indeed, the photo-z accuracy is not crucial for the
simulation. The COSMOS catalog is only used to obtain a representative
population of galaxies in term of density and mix of galaxy types.
Since the predicted apparent magnitudes are calculated from the
best-fit templates, the photo-z accuracy has no impact on our ability
to link the simulated redshift to the predicted colors. The only risk
of including lower quality photo-z's is to degrade slightly the
catalog representativity, possibly biasing the redshift distribution
at $z>1.25$.

\subsubsection{The mock catalog construction}

The principle of our simulation is to convert the observed properties
of each COSMOS galaxy into simulated properties that can then be
viewed using any possible instrument configuration. A photo-z and a
best-fit template (including possible additional extinction) are
associated with each galaxy of the COSMOS catalog.

The first step is to integrate the best-fit template (in the observer
frame) through the instrument filter transmission curves to produce
simulated magnitudes in the instrument filter set.

The second step is to apply random errors to the simulated magnitudes,
based on the magnitude-error relations established in each filter (see
Section \ref{subsubsec:noise}).

Importantly, all the COSMOS measured properties are propagated to the
simulated galaxies, for instance, the galaxy half-light radius as
measured on the ACS images by \citet{Leauthaud07}.

This approach presents the following advantages:
\begin{itemize}
\item the simulated mix of galaxy populations is, by construction,
representative of a real galaxy survey,
\item additional quantities measured in COSMOS (like the galaxy size,
UV luminosity, morphology, stellar masses, correlation in position)
can be easily propagated to the simulated catalog.
\end{itemize}

The COSMOS mock catalog is limited to the range of magnitude space
where the COSMOS imaging is complete ($i^+_{AB}\sim 26.2$ for a
$5\sigma$ detection, see \citet{Capak07} and Capak et al. (2009) in
prep.

\subsubsection{Simulating emission lines}
For each galaxy of the COSMOS mock catalog we have associated emission
line fluxes. This feature is useful to predict the size and the depth
of a spectroscopic redshift sample. We modeled the emission line
fluxes (Ly$\alpha$, [OII], H$\beta$, [OIII] and H$\alpha$) of each
galaxy as explained below.

We used the method described in Section 3.2 of \citet{Ilbert09}.
Using the \citet{Kennicutt98} calibration, we first estimated the star
formation rate (SFR) from the dust-corrected UV rest-frame luminosity
already measured for each COSMOS galaxy. The SFR can then be
translated to an [OII] emission line flux using another calibration
from \citet{Kennicutt98}. We checked that the relation found between
the [OII] fluxes and the UV luminosity is in good agreement with the
VVDS data (see Fig.3 of \citet{Ilbert09}) and still valid for
different galaxy populations. For the other emission lines, we adopted
intrinsic, unextincted flux ratios of [OIII]/[OII] = 0.36;
H$\beta$/[OII] = 0.28; H$\alpha$/[OII] = 1.77 and Ly$\alpha$/[OII] = 2
(\citet{McCall85}, \citet{Moustakas06}, \citet{Mouhcine05},
\citet{Kennicutt98}). The approach of determining the Ly$\alpha$ line
flux through its ratio with the OII emission line flux is perhaps not
good, but since the Ly$\alpha$ line becomes visible to
optical-NIR wavelength surveys only beyond $z\sim 3$, it will not have a big
impact.
Finally, we reduce each galaxy's line flux using the best-fit dust
attenuation found with the template fitting procedure in the COSMOS
photo-z catalog.

The same procedure can be applied to the GLFC since we can calculate
the UV absolute luminosity the same way as for the CMC.

The COSMOS mock Catalog (CMC) has the advantage over the GLFC that it
better preserves the relations between galaxy size and color (and
presumably type and redshift). The CMC may also reproduce color
distributions more accurately, since the population has not been
reduced to three galaxy types as in the GLFC. The CMC is, however,
limited by construction to the range of magnitude space where the
COSMOS imaging is complete, whereas the GLFC can be extrapolated to
fainter galaxies. Our validation tests below will compare the
differences of these two simulation approaches. It is important to
stress that all the following galaxy number densities correspond to
mask-corrected areas. Hence for real surveys, those numbers would have
to be reduced by a factor of $\approx10\%$ as observed in COSMOS
field.

\section{Validation}
\label{sec:validate}
The aim of these mock catalogs is to predict the performance of future
surveys such as JDEM and EUCLID. The depth of the foreseen surveys
may be significantly fainter than the deepest existing spectroscopic
and NIR imaging wide field surveys, so a comprehensive observational
validation of the catalog is not yet possible---especially in terms of
NIR photometry and redshift distribution. We should, however, prove
that the GLFC and CMC are consistent with existing data. For this
validation, we will compare the galaxy counts, color distributions,
redshift distributions, and emission-line distributions to a selection
of the deepest available relevant survey data. The comparison data are
taken from:

\begin{itemize}
\item The HST Ultra-Deep Field (UDF) catalog \citep{Coe06}
\footnote{http://adcam.pha.jhu.edu/$\sim$coe/UDF/} and included
references covers 11.97 arcmin$^2$ in the 4 HST/ACS filters F435W,
F606W, F775W and F850LP, referred to as $B$, $V$, $i$, and $z$ bands,
respectively, and 5.76 arcmin$^2$ for the NICMOS filters F110W and
F160W, referred to as J and H bands. The UDF detects objects are at
$10 \sigma$ for $z<28.43$, $i<29.01$ and $J<28.3$, significantly
fainter than the expected depth of any of the aforementioned proposed
surveys. Due to the small area, the UDF is more sensitive to the
cosmic variance, but it is useful for comparisons at faint magnitudes.
\item The GOODS (Great Observatories Origins Deep surveys) v1.1 survey
  catalog \footnote{http://archive.stsci.edu/prepds/goods/}
  \citep{Giavalisco04}. The GOODS observations are split into northern
  and southern fields. Each field covers 160~arcmin$^2$ in the same
  $BViz$ filters as the UDF, and is $90\%$ complete at $z\approx26$.
\item The GOODS-MUSIC (MUltiwavelength Southern Infrared Catalog)
  sample \citep{Grazian06} has been constructed with public data of
  the GOODS-S field, NIR Spitzer data from IRAC instrument (3.6, 4.5,
  5.8 and 8.0 $\mu$ m) and U-band data from the 2.2ESO and VLT-VIMOS
  covering 140~arcmin$^2$. This catalog is both z and Ks-selected and
  is $90\%$ complete for $K_s<23.8$ and $z<26$.
\item The VVDS-DEEP first epoch public
  release\footnote{http://cencos.oamp.fr/} includes photometric data
  in BVRI VIRMOS-VLT bands over 0.49~deg$^2$ \citep{McCracken03}; a
  spectroscopic survey of targets with $I_{AB}<24$ \citep{LeFevre05}
  with a sampling rate of 0.2 and a mean redshift of 0.86.
\item The VVDS-DEEP NIR $J$ and $K_s$ photometry \citep{Iovino05}
  covers 170~arcmin$^2$ and is complete for $Ks<22.5$. This sample
  contains the $BVri$ VIRMOS-VLT band and the $JKs$ ESO/NTT (New
  Techonology Telescope) using the SOFI Near Infrared imaging camera.
\end{itemize}
A cut in the size-magnitude plane is applied to each dataset in order
to remove stellar contamination, in case this has not already been
done for published catalogs. \\ For each of the above listed surveys,
both a CMC and a GLFC are constructed as described above, projecting
both mock catalogs onto the real surveys' filter sets.

\subsection{Galaxy Counts}
\begin{figure}[!ht]
\resizebox{\hsize}{!}{\includegraphics{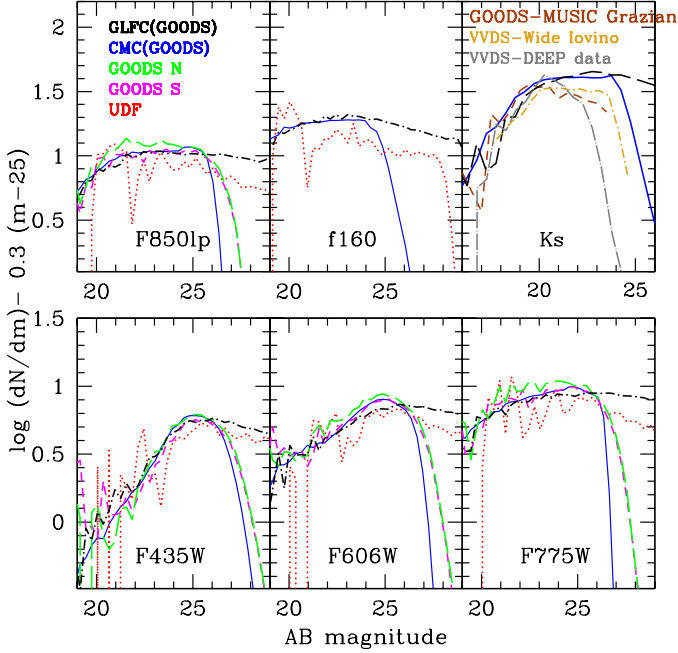}}
\caption{Differential galaxy counts for F435W ($B$ band), F775W ($i$
  band), F850LP ($z$ band), F160W ($H$ band) and the Ks band comparing
  mock catalogs to real observations of UDF and GOODS surveys.}
\label{counts}
\end{figure}
We compare the simulated galaxy counts ($dN/dm$) to those in
the real catalogs listed above. The SED of each simulated galaxy is
integrated over the bandpass of each real catalog.

In {\bf Figure~\ref{counts}} we compare the differential galaxy counts
in the $B, V, i$, and $z$ bands of the UDF and GOODS to those
synthesized from the GLFC and CMC simulations. The blue solid line is
the COSMOS mock catalog (CMC) and the black dot-dashed line is the
{\it Le Phare simulation} based on the GOODS LF (GLFC). The real
observations are: the UDF (red dotted); GOODS North (long-dashed
green) and South (dashed magenta). The UDF counts in the NIR are also
compared to the simulations. An excellent agreement is seen between
all sources for $21<m<26$, where each is complete, with some tendency
for the UDF counts to be lower than other surveys. We attribute this
to cosmic variance because of the very small UDF survey area. The
difference between the simulated catalogs and the GOODS South is
generally less than the difference between GOODS North and South
($\approx15\%$). Note that the CMC becomes incomplete for $m>26$
($m>25$ in the 1.6~$\mu$m band), so will underestimate the galaxy
yield in surveys deeper than these limits.

We also compare in Figure~\ref{counts}, the simulated $Ks$ band counts
to the GOODS-MUSIC and VVDS data. The $K_s$-band VVDS data of
\citet{Iovino05} are $90\%$ complete at $I_{AB}=25.5$. We see a very
good agreement between the VVDS, GOODS-MUSIC and simulated catalogs at
$Ks<20$ and some discrepancies at fainter magnitudes. However, we
measure a mean less than $17.9\%$ difference between VVDS Iovino and
CMC, $14.4\%$ difference between VVDS Iovino and GLFC and $17.4\%$
between GOODS-MUSIC and VVDS Iovino within the observation limits of
these 2 surveys. We conclude that both simulated catalogs are well
reproducing the Ks counts.

\subsection{Colors}
\begin{figure}[!ht]
\resizebox{\hsize}{!}{\includegraphics{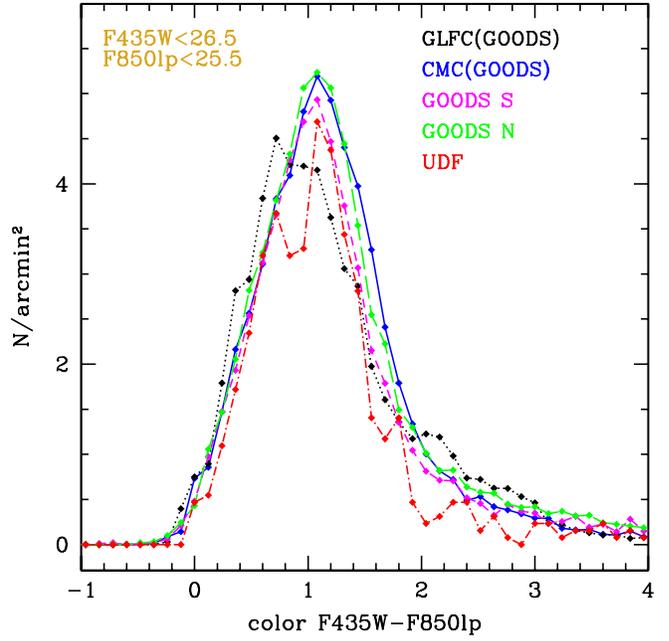}}
\caption{Color histogram of F435W ($B$ Band)-F850lp ($z$ Band)
  comparing mock catalogs to real observation of UDF and GOODS surveys.}
\label{colorsBz}
\end{figure}
{\bf Figure~\ref{colorsBz}} compares the $B-z$ color distribution of
the simulated catalogs to those of GOODS and UDF galaxies. The blue
solid line is the COSMOS catalog (CMC) and the black dotted line is
the GLFC catalog in GOODS filters. The real observations are: the UDF
(dot-dashed red); GOODS North (long-dashed green) and South (dashed
magenta). The indicated cut in S/N and magnitude are applied to F850LP
($z$ Band) for each catalog. The magnitude cuts in the B and z bands
are taken following the COSMOS completness in each of these
bands. There is a good agreement in these optical wavelengths. The
simulated catalogs have median and mean $B-z$ colors in agreement at a
few percent with the UDF and GOODS catalogs (see {\bf
  Table~\ref{meantab}}).

The UDF seems relatively deficient in the reddest galaxies, again
perhaps a manifestation of cosmic variance, which is most severe for
the highly-clustered red-sequence galaxies. This is an expected result
and is not an issue with the catalogues since the UDF may be
under-dense due to its small survey area.

\begin{table}[!ht]
\caption{Mean and median of the $B-z$ color distributions with the
  magnitude cuts of {\bf Figure~\ref{colorsBz}} corresponding to the
  completeness of the CMC.}
\begin{tabular}{cccccccc} \hline\hline
Catalogs & Mean & Median & survey area & mag limit \\
CMC(GOODS) & 1.24 & 1.14 & 1.24 deg$^2$ & I $\sim$ 26 \\
GLFC(GOODS) & 1.2 & 1.05 & 0.1 deg$^2$ & R $\sim$ 31 \\
GOODS-S & 1.26 & 1.1 & 160 arcmin$^2$ & z $\sim$ 26 \\
GOODS-N & 1.3 & 1.13 & 160 arcmin$^2$ & z $\sim$ 26 \\
UDF & 1.16 & 1.06 & 11.97 arcmin$^2$ & z $\sim$ 28.43 \\
\hline\hline
\label{meantab}
\end{tabular}
\end{table}

\begin{figure}[!ht]
\resizebox{\hsize}{!}{\includegraphics{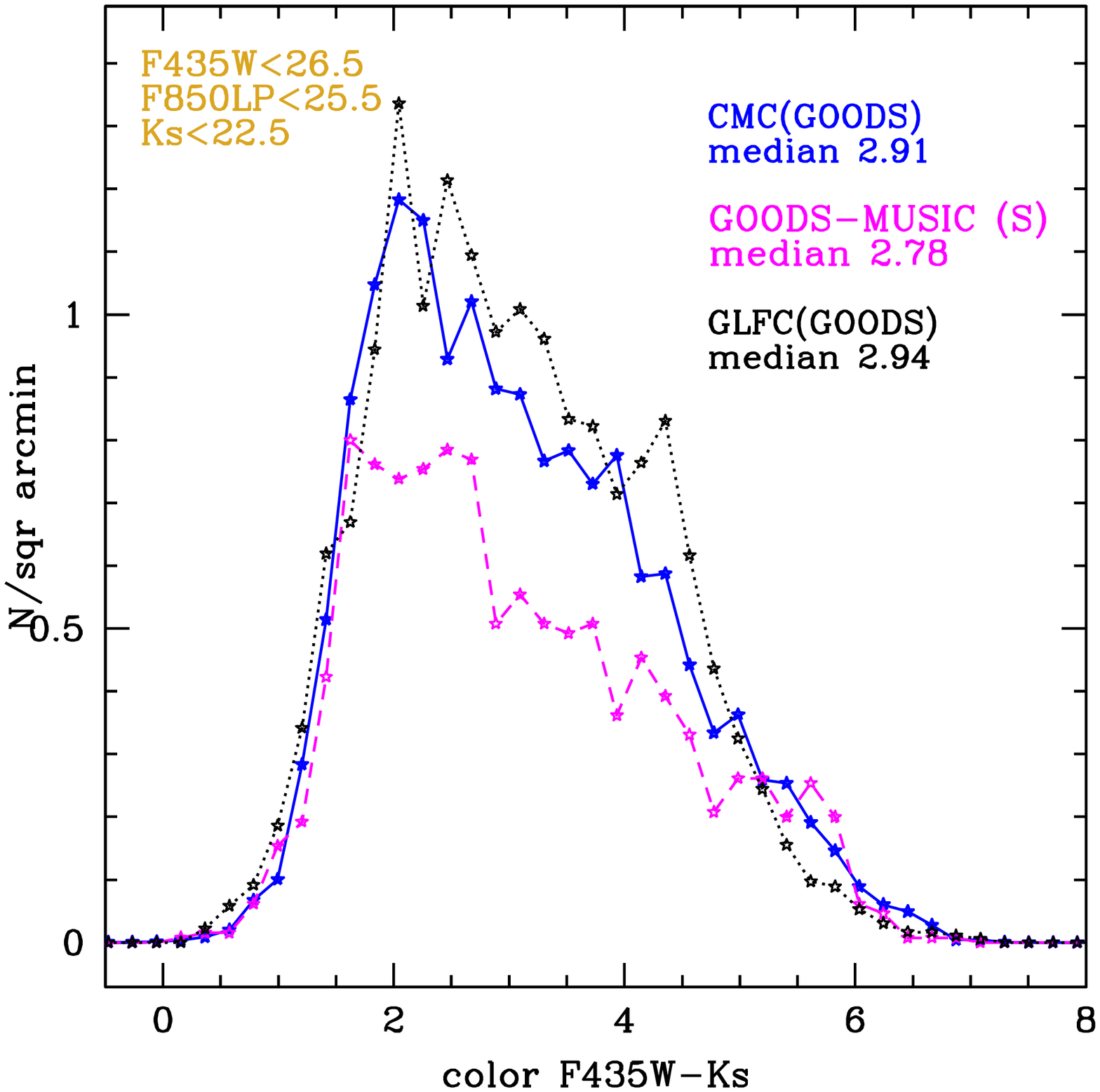}}
\resizebox{\hsize}{!}{\includegraphics{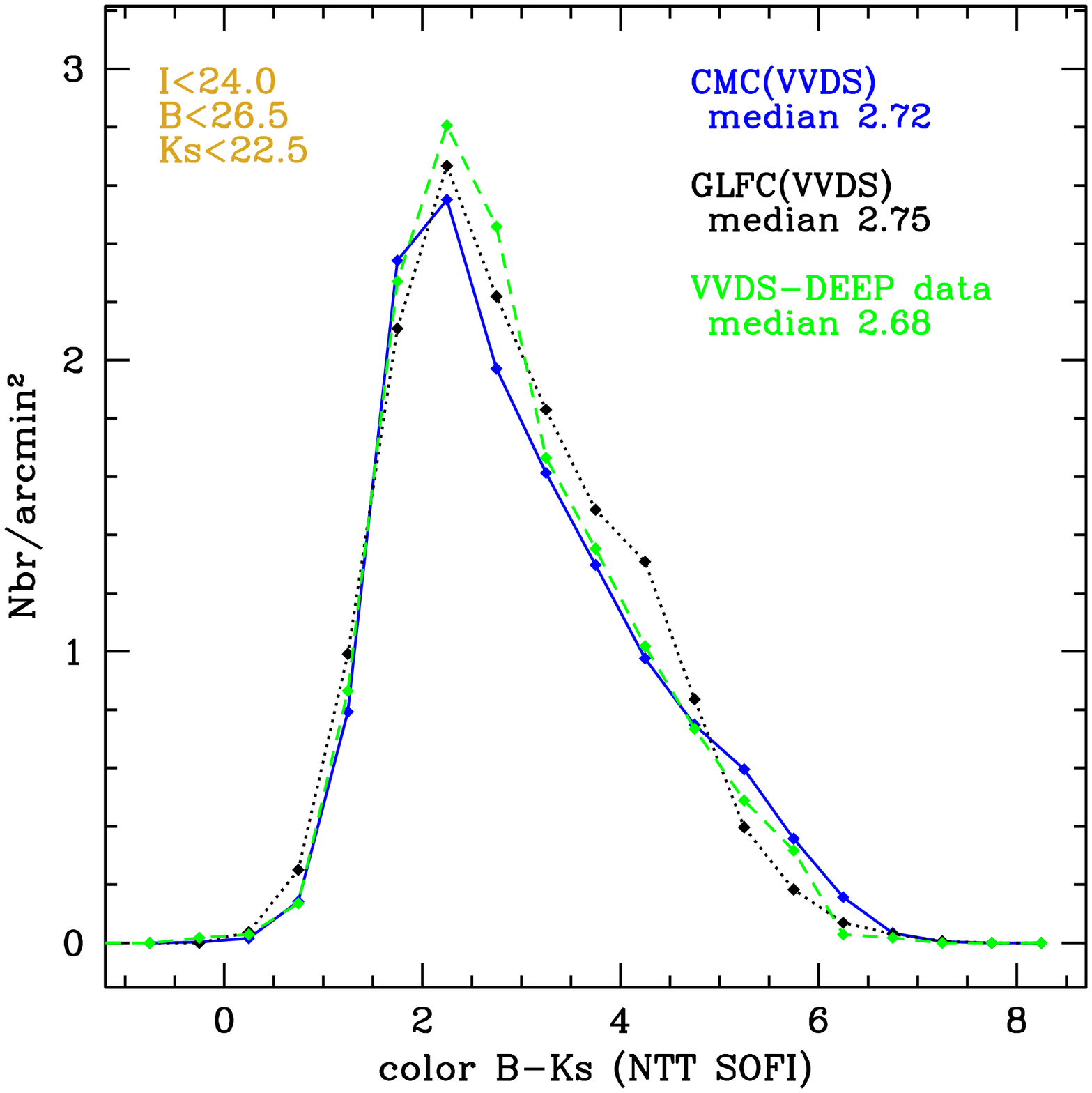}}
\caption{Color histogram of F435W ($B$ Band)-Ks comparing mock
  catalogs to real observation of GOODS and VVDS-DEEP surveys.}
\label{colorsBK}
\end{figure}

{\bf Figure~\ref{colorsBK}} compares simulated $B-Ks$ colors to those
in the GOODS-MUSIC and VVDS-DEEP catalog.  The blue solid line is the
COSMOS mock catalog (CMC) and the black dotted line is the GLFC
catalog. The observational data are the GOODS-MUSIC survey (top panel
in dashed magenta) and VVDS-DEEP survey (bottom panel in dashed
green). We choose to cut at 22.5 AB mag in the $Ks$ Band due to the
VVDS incompletness and the variability of the GOODS-MUSIC survey area
beyond this magnitude. The color distributions in NIR agree well
inside the completeness limits imposed by the different surveys, the
CMC $z<25.5$ and $B<26.5$ and the GOODS-MUSIC survey $Ks<22.5$. The
GOODS-MUSIC data have lower B-Ks counts which is explained by the
number count difference at $Ks>20$ in the Ks-band between both
simulated catalogs and the Grazian counts (see
Figure~\ref{counts}). However, the mean and median colors are in good
agreement. The agreement is much better when comparing our mock
catalogs to the VVDS-DEEP survey (bottom panel of
Figure~\ref{colorsBK}).

\subsection{Redshifts}
\begin{figure}[!ht]
\resizebox{\hsize}{!}{\includegraphics{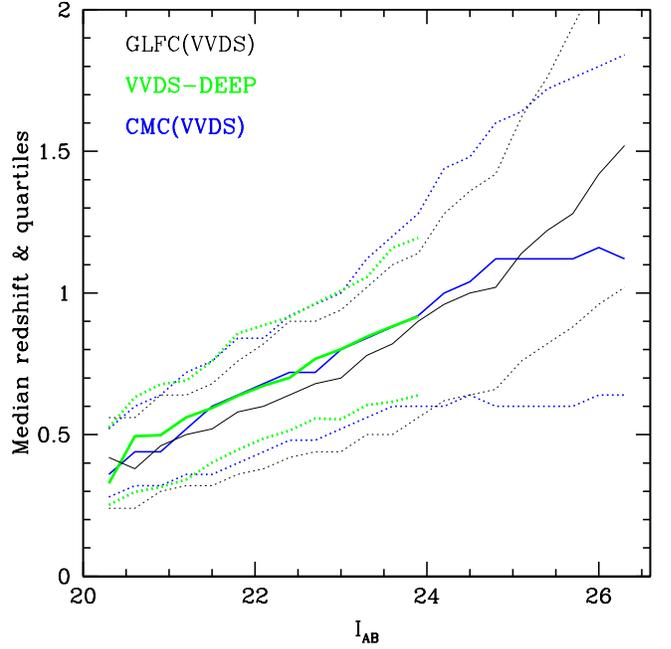}}
\caption{Redshift distribution as a function of the VVDS $i$
band magnitude compared to mock catalogs redshift distribution.}
\label{zmed}
\end{figure}
\begin{figure}[!ht]
\resizebox{\hsize}{!}{\includegraphics{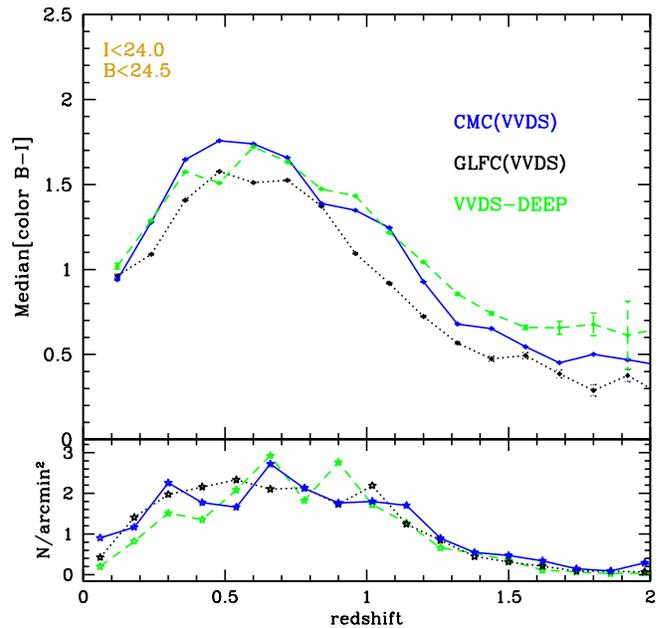}}
\caption{Median of B-i color (VVDS bands) as a function of redshift
  for GLFC and CMC mock catalogs and the VVDS-DEEP survey.}
\label{zcol}
\end{figure}
\begin{figure}[!ht]
\resizebox{\hsize}{!}{\includegraphics{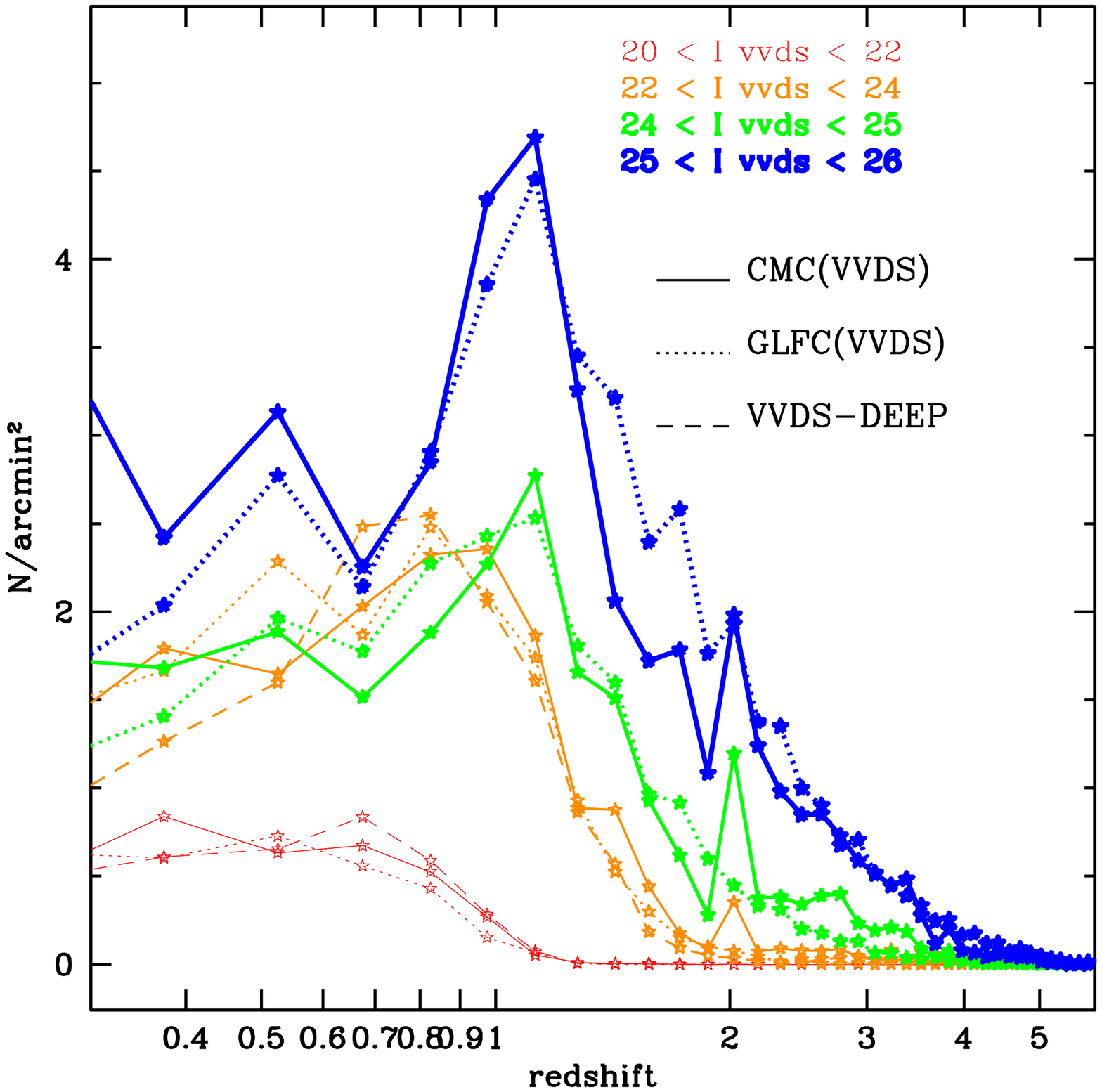}}
\caption{Redshift distribution of the VVDS-DEEP survey, the COSMOS
  survey and the GLFC catalog from the GOODS luminosity function.}
\label{redshift}
\end{figure}

Comparison of the simulated redshift distributions to real data is
likely less accurate than colors or magnitudes, since real redshift
surveys are much shallower and have significant incompleteness.
Nonetheless, {\bf Figure~\ref{zmed}} shows median and quartiles of the
redshift distributions vs $I$-band magnitude of the CMC (blue
medium-thickness line) and the GLFC catalogs (black thin line) in the
VVDS filters compared to the VVDS-DEEP redshift distribution (green
high-thick line). The dotted lines represents the quartiles and solid
lines the median of the redshift distribution. 

The redshift quartiles of the CMC agree very well with the VVDS
spectroscopic redshift distribution to the $I=24$ limit of the latter.
The GLFC seems to have a lower median redshift, probably due to the
GOODS LF used to produce the catalog. The mean difference of the
median redshift is 0.02 and 0.07 for the CMC and GLFC catalogs,
respectively. We conclude not surprisingly that the CMC is probably a
better representation of the magnitude-redshift distribution, and is
as accurate as current data can validate.  

{\bf Figure~\ref{zcol}} plots median $B-i$ color vs redshift for the
simulated catalogs CMC (blue solid line) and GLFC (black dotted line)
in the VVDS passbands compared to the VVDS-DEEP survey (green dashed
line). All catalogs have been restricted to $I<24$, the completeness
domain of the VVDS-DEEP redshift survey. The colors agree to 0.1--0.2
mag until $z\sim2$. Above this redshift the VVDS survey is likely to
be highly incomplete, even for blue galaxies, as no strong emission
lines are available in the VVDS spectral wavelength range and due to
the $I=24$ magnitude cut (see Figure~\ref{zmed}) for the median
redshift and quartiles of the distribution.

We compare the redshift distributions ({\bf Figure \ref{redshift}})
for different magnitude cuts (different colors and thickness). There
is a very good agreement beetween the VVDS-DEEP (dashed lines), the
COSMOS survey (solid lines) and the GLFC (dotted lines) redshift
distributions.

The CMC and the VVDS-DEEP catalog show very good agreement. This
figure shows that both mock catalogs are well reproducing the number
density inside the VVDS volume : ($0<z<1.5$ and $I<24$) and agree at
higher magnitude and redshift. The GLFC is based on the GOODS
luminosity function (\citep{Dahlen05}) which has equivalent depth to the
VVDS galaxy sample. However when looking at Figure \ref{zmed}, the
VVDS magnitude-redshift distribution seems to be in better agreement with
the CMC than the GLFC.

\subsection{Emission-Line Strength}
\begin{figure}[!ht]
\resizebox{\hsize}{!}{\includegraphics{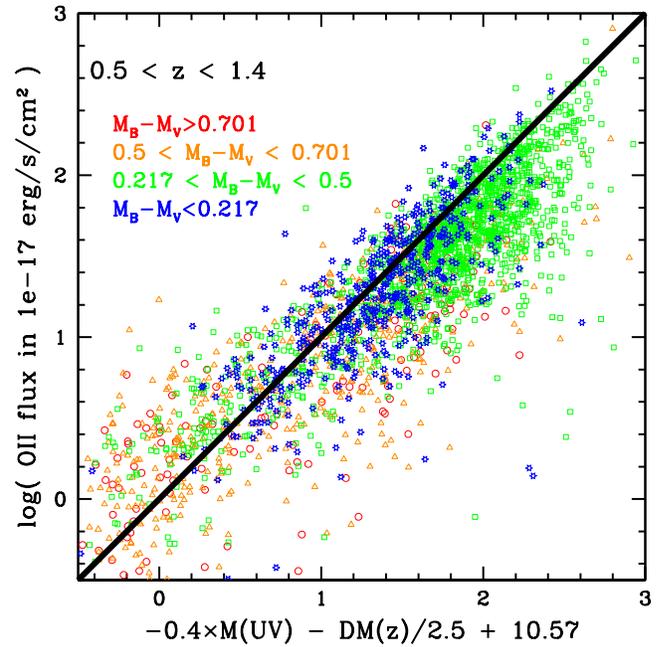}}
\caption{UV-OII correlation using 4 color cuts representing the early,
  intermediate, late and starburst galaxies using the VVDS data from
  \citet{Lamareille08}.}
\label{uvo2}
\end{figure}

\begin{figure}[!ht]
\resizebox{\hsize}{!}{\includegraphics{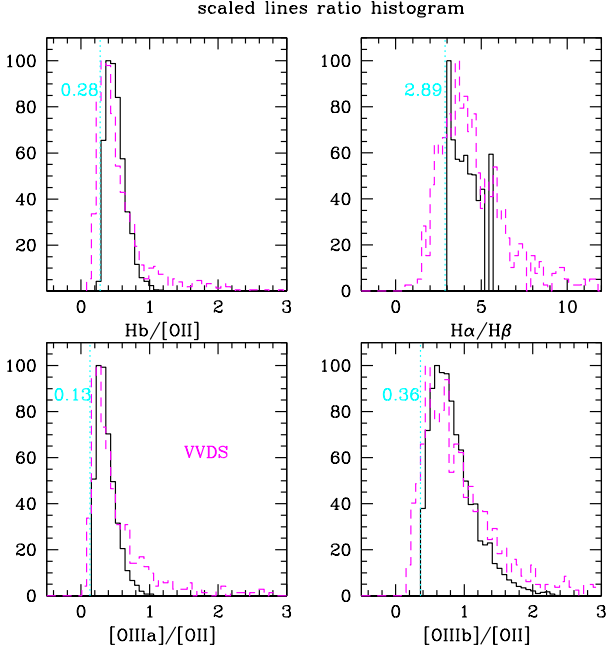}}
\caption{Emission line ratios for the CMC catalog in black solid line
  compared to VVDS ratios in magenta dashed line. The cyan dotted line
  is the value of the theoretical emission line ratios.}
\label{eml}
\end{figure}

Future dark energy surveys need large spectroscopic redshift
samples for calibrating photometric redshifts, measuring accurately
the BAO, and for other probes using spectroscopic samples of
galaxies. Thus it is crucial to have realistic emission lines
allowing predictions of the success rate, depth, and size of
spectroscopic samples that we will need. 
Although, the CMC does not reproduce absorption lines, it
simulates emission lines for all galaxies in the catalogue, which
allows a first estimation of spectroscopic survey capabilities. The
validation of the emission line strength can be found in Figure 3 of
\citet{Ilbert09}. This figure shows the relation between the OII flux
and the rest-frame UV luminosity predicted by \citet{Kennicutt98} :
\begin{equation}
log[OII]=-0.4M_{UV} + 10.57 -\frac{DM(z)}{2.5} 
\end{equation}
It shows very good agreement between VVDS data from
\citet{Lamareille08} and the simulated emission lines strength
extrapolated from the photometric redshift best fit template between
$0.4<z<1.4$ where the OII emission line can be measured. Figure
\ref{uvo2} shows the UV-OII relation using the VVDS data from
\citet{Lamareille08} with 4 color cuts corresponding to the early
types in red circles, intermediate types in orange triangle, late
types in square green and starburst galaxies in blue stars. Criteria
for the type selections follows the Dalhen prescription detailed in
the Table \ref{BVrest}. This shows that the correlation has no strong
dependence on the galaxy types.
The histograms shown in Figure~\ref{eml} are the emission line ratios
of the CMC compared with those of VVDS. The attenuation causes a
spread on the emission line ratios but we had to add a Gaussian
dispersion to fit the VVDS emission line spread. If $r_{n}$ is a set
of $n$ numbers randomly drawn from a Gaussian distribution
$(\mu,\sigma^2)=(0,1)$, we define the new fluxes as :\\
\begin{eqnarray}
&H\alpha& = H\alpha(1+\left|r_{0}/4\right|) \\
&H\beta& = H\beta(1+\left|r_{0}/2\right|) \\
&[OIII]_{4959}& = [OIII]_{4959}(1+\left|1.2r_{1}\right|) \\
&[OIII]_{5007}& = [OIII]_{5007}(1+\left|r_{2}\right|)
\end{eqnarray}
We use the same random number $r_{0}$ to spread H$\alpha$ and
H$\beta$ in order to preserve the ratio of these lines. \\ The
spectral wavelength range of VVDS spans from 0.55 to 0.94$\mu$m
\citep{LeFevre05} making the $[OII]$ line visible from redshift 0.5 to
1.5 and the $H\alpha$ line visible from redshift 0 to 0.5. Thus we
choose to evaluate the validity of the $H\alpha$ fluxes using the
ratio $H\alpha/H\beta$, known as the Balmer decrement and having a
value $\sim2.9$. The spread of the CMC emission line ratios fits well
the emission line spread observed in the VVDS data.

\begin{figure}[!hb]
\resizebox{\hsize}{!}{\includegraphics{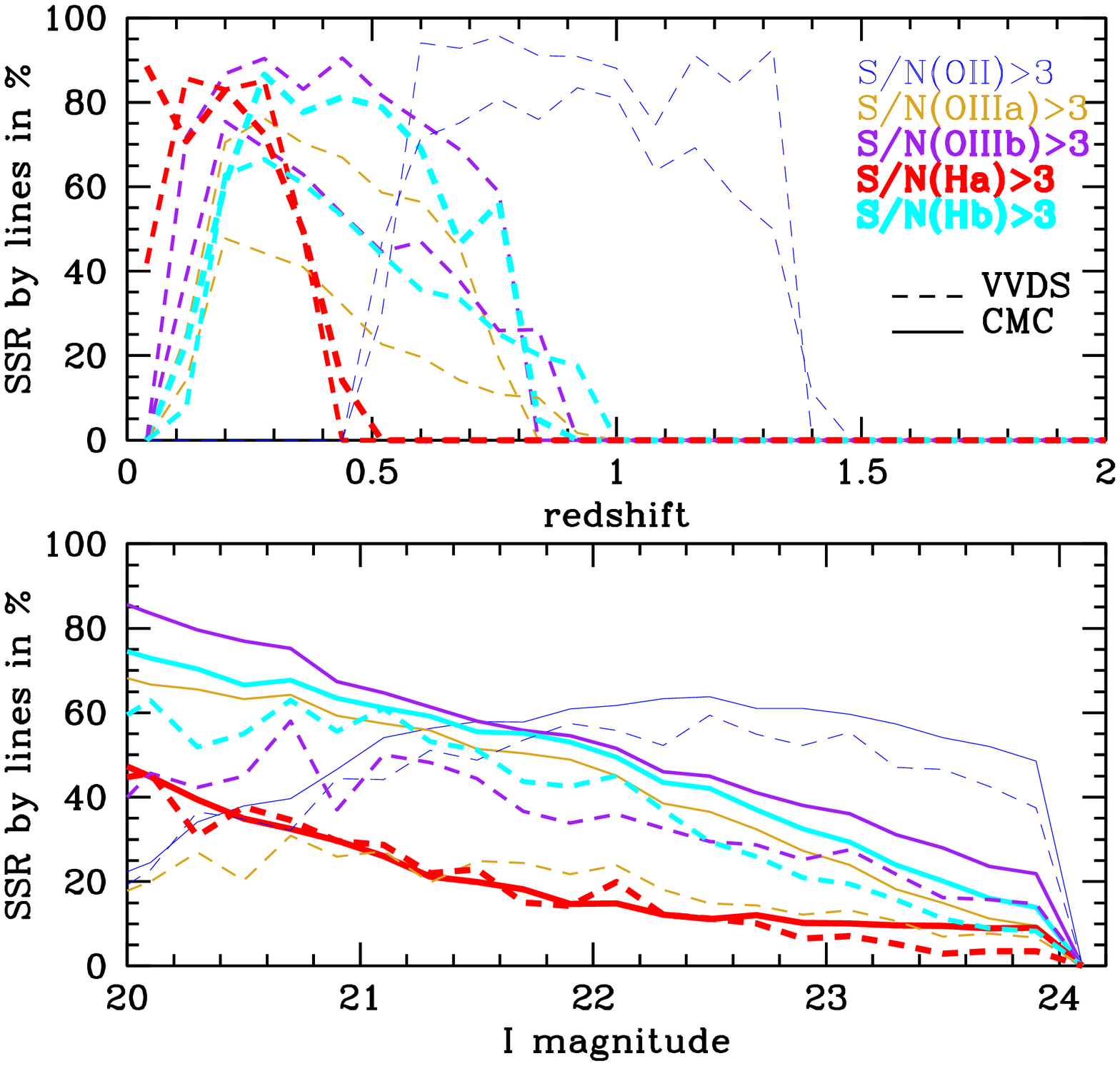}}
\caption{Comparison of the lines Spectroscopic Success Rate (SSR) of
the VVDS survey with the simulated lines VVDS SSR using the CMC
emission lines. }
\label{ssrlines}
\end{figure}
\begin{figure}[!hb]
\resizebox{\hsize}{!}{\includegraphics{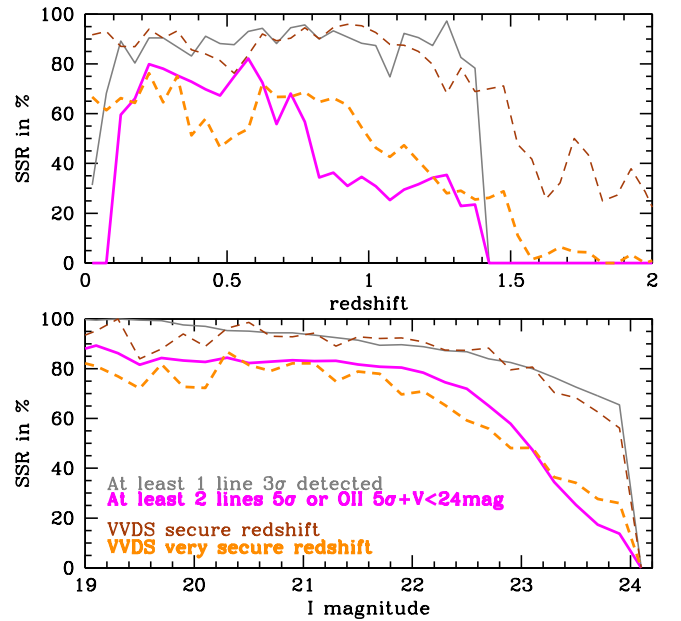}}
\caption{Comparison of the Spectroscopic Success Rate (SSR) of the
VVDS survey with the simulated VVDS SSR using the CMC emission
lines.}
\label{ssr}
\end{figure}
For a further validation of the CMC emission lines we choose to
reproduce the spectroscopic success rate (SSR) of the VVDS survey. To
realise the CMC(VVDS) SSR we have extrapolated the flux sensitivity of
the VVDS survey as a function of wavelength for several
signal-to-noise ratio based on the [OII] emission lines. Using these
sensitivities and the CMC emission line catalog we derive the
simulated SSR for the VVDS survey in order to validate the emission
line model of the CMC catalog (Figures~\ref{ssrlines} and~\ref{ssr})
. {\bf Figures~\ref{ssrlines}} show the SSR for the emission lines of
the CMC catalog compared to those of the VVDS as a function of
redshift and magnitude. The top panel shows the SSR of emission lines
as a function of redshift, dashed lines for the VVDS survey and solid
lines for the CMC. The blue lines are for the OII lines, red for
H$\alpha$, cyan for H$\beta$, gold for OIIIa at $4959\AA$ and purple
for OIII at $5007\AA$ using a growing thickness for each one in that
order. The bottom panel shows the same as the top panel as a function
of magnitude. The strong emission lines, H$\alpha$ and OII, are in
very good agreement with the VVDS both in magnitude and redshift
space.  However weaker emission lines have a discrepancy of 10 to 20
$\%$ compared to the VVDS emission lines at low magnitudes, especially
for the [OIII] lines at $4959\AA$ and $5007\AA$, but it agrees very
well at magitudes $I>21$. {\bf Figures~\ref{ssr}} compare the overall
SSR predictions of the CMC with the VVDS secure and very secure
redshifts. The top panel shows the VVDS SSR for secure redshift (brown
thin lines) and very secure redshift (orange thick line) as a function
of redshift compared to the CMC-VVDS emission lines $3\sigma$ flux
detection (grey thin lines) and 2 lines at $5\sigma$ detection or OII
at $5\sigma$ detection with $V<24$ (magenta thick lines). The bottom
panel shows the same as the bottom panel as a function of magnitude.
We see that the VVDS secure redshifts success rate is very close to
the 3$\sigma$ detection line of the CMC catalog. In the same way, the
VVDS very secure redshift success rate agrees very well with the 2
lines detection at 5$\sigma$. This includes galaxies with $V<24$,
which corresponds to a detection of the continuum in the VVDS spectra
with a $S/N>10$ in the blue part of the spectrum (which is free of
strong sky emission lines). The differences mainly comes from the
redshifts effectively obtained using absorption lines for which the
shape of the continuum altogether does not match exactly with our
ad-hoc V-band magnitude criterion. Although the CMC does not simulate
the absorption lines, we can say that the CMC emission line shows a
good agreement both in redshift and magnitude with the VVDS
spectroscopic survey. These results makes us confident in using the
CMC catalog as a tool to predict the SSR of future wide field
spectroscopic surveys.

\section{Discussion}

Different cosmological tests have been proposed to probe the geometry 
and growth of structures in order to shed new light on 
the nature of dark energy. The best approach is certainly to combine 
different probes to reduce the errors and better understand the systematics. However, each 
probe has its own requirements and it is a technical and scientific 
challenge to design a telescope optimised for more than one cosmological probe.
Nonetheless, the goal of future dark energy programs should find a survey strategies that will lead to the best combined efficiency of different probes. 

Having a realistic description of galaxy properties in our Universe 
is key to properly forecast what future deep and wide surveys can achieve 
for a given survey configuration. In this section, we start a discussion 
with requirements for a dark energy survey that combines
shape measurements and a photo-z calibration
survey (PZCS) for weak lensing with a BAO survey.

Thus, we will discuss two important aspects of these future dark
energy mission: {\it (i)} the impact of galaxy size in terms of shape
measurement for weak lensing depending on the PSF size {\it(ii)} the
different typical WL and BAO survey configurations we have tested using
our mock catalogues and {\it (iii)} their expected spectroscopic
success rate.

Note that we are not conducting here any optimisation of such a space
mission and its survey strategy. We are just putting in place some of
tools necessary for such an optimisation.

\subsection{Galaxy-sizes}
\begin{figure}[!ht]
\resizebox{\hsize}{!}{\includegraphics{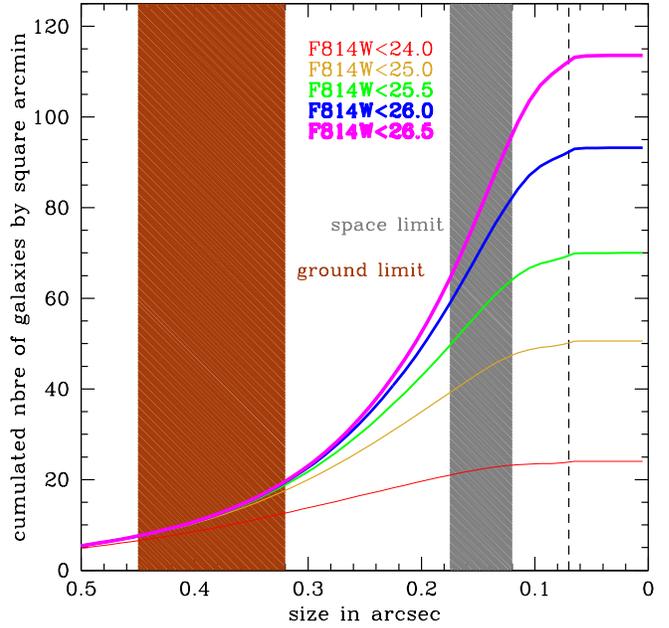}}
\caption{Comparison of a ground-based and space weak-lensing survey
  using the COSMOS cumulated half-light radius distibution by bins of
  magnitude.}
\label{size}
\end{figure}

A comparison of the simulated galaxy-size distribution to
observational data is not particularly useful. Indeed the simulated
catalogs agree, by construction, with the size distribution measured
in the COSMOS ACS imaging survey \citep{Koekemoer07},
\citep{Leauthaud07}. No other survey capable of resolving faint
galaxies approaches the size of the COSMOS/ACS data. In {\bf
  Figure~\ref{size}} we show the cumulated half-light radius
distribution for the COSMOS data by bins of magnitude (different
colors and thickness).

The brown area corresponds to the typical PSF size of ground-based
telescopes. It shows that we can only resolve and measure galaxy
shapes (those galaxies with a size larger than the PSF size)
for about $15$gal/arcmin$^2$ (or up to
$20$gal/arcmin$^2$ in case of excellent seeing) as soon as a depth of
$I\sim25$ is reached. Going deeper than $I=25$ will not help to raise
this number density as fainter galaxies are smaller than the PSF, thus
making a shape measurement extremely difficult.

The grey area corresponds to the PSF size expected with future space
Dark Energy mission; the exact value of the PSF size depends on the
telescope diameter and the chosen pixel scale (here we assume a
telescope diameter of $\sim 1.5-1.8$ meter). Interestingly, with this
small PSF size, the number density of resolved galaxies does increase
as a function of the depth of the survey. Beyond $I=26$ the increase
in number density is however smaller, suggesting that there is a
limited gain to go much deeper than $I=26$ except perhaps for telescope
designs with the smallest PSF size. With a depth of $I\sim 25.5-26$
the number density of resolved galaxies in the foreseen space surveys
is about 3$\times$ larger than what ground-survey can achieve.

Contrary to the ground case, the performance of a space survey in
terms of galaxy number density critically depends both the depth and the resolution
of the images. The black dashed
vertical line correspond to the PSF size of the ACS/HST camera as
measured in the COSMOS survey \citet{Leauthaud07}. The COSMOS survey
reach a number density of resolved galaxies of $\sim
110$gal/arcmin$^2$ down to $I\sim26.5$. It is important to recognize
however that not all these galaxies could be used in the COSMOS 3D
weak lensing analysis, as explained in Leauthaud et al (2007). For a
fraction of these galaxies there was no counterpart in the
ground-based catalogue because of masking of the data. For the
remaining galaxies, it was not always possible to determine securely
the shape or the redshift thus decreasing further the number of usable
galaxies to roughly 40gal/arcmin$^2$.

Cosmological weak lensing survey aim to maximize the number of
resolved galaxies that can be used for weak lensing tomography. As
shown by \citet{Amara07}, it is more efficient to conduct a wider
survey than a deeper survey as the galaxy number increase is larger
for a given exposure time by going wide than by going deep. However,
the increase of Galactic absorption and stellar density, when getting
closer to the Milky Way plane, will limit the gain of going wider than
deeper when surveying areas larger than $\sim$15 000 sq.deg. 


\subsection{BAO and WL spectroscopic survey requirements}
\begin{figure}[!ht]
\resizebox{\hsize}{!}{\includegraphics{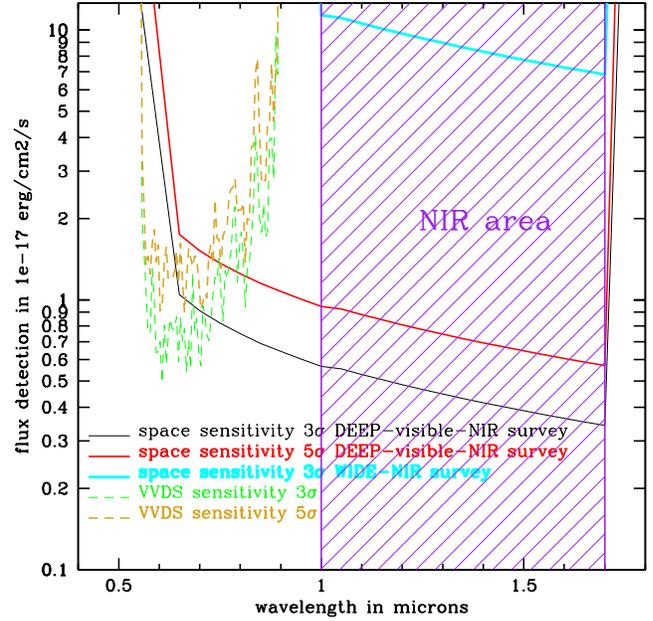}}
\caption{Flux sensitivities for the VVDS-DEEP survey as a function of
  wavelength compared to forecast of future space WIDE and DEEP NIR
  [purple and shaded area] surveys with or without visible coverage.}
\label{jdemsens}
\end{figure}

We investigate here three types of spectroscopic redshift surveys, as
can be seen in {\bf Figure~\ref{jdemsens} and Table~\ref{surveys})},
whose characteristics correspond to typical requirements for the BAO
and WL probes. \\

{\it (i)} BAO aims to measure the baryon accoustic oscillation peak of
the 2-pt correlation function as a measure of a standard ruler.  It
requires a large survey area to reduce the statistical noise
\citep{Glazebrook05,Blake03} and an accurate redshift for each galaxy,
typically using spectroscopic techniques to reach a higher
accuracy. One can use photometric redshifts instead of spectroscopic
redshifts, but the loss of line-of-sight information then requires a
few times larger area to reach the same dark energy figure of merit
(FOM) \citet{Glazebrook05}. Following this last paper, we investigate
a WIDE near infrared (1.0 to 1.7 micron) spectroscopic survey reaching
a 3$\sigma$ sensitivity of $1x10^{-16}ergs\ cm^{-2} s^{-1}$ at
1.2$\mu$m (this can be achieved with an efficient slitless
spectrograph using a 1.5m telescope, 0.28''/pixel, R$\sim$500 and an
exposure time of 1200 sec) and covering a large area on the sky to
maximise the dark energy FOM. The flux sensitivities of such a survey
are represented by the solid cyan (and high thickness) line in {\bf
  Figure~\ref{jdemsens}}.


{\it (ii)} The weak-lensing tomographic analysis is likely the most
efficient lensing method to estimate the dark energy parameters
\citep{Massey07,Amara07}. This method probes the growth of structures
but needs to have accurate redshifts to place the background sources
at their correct location.  To achieve this measurement, it is
essential to determine with the best accuracy (and minimal biases) the
photometric redshift (photo-z) of all the galaxies to be used in the
weak lensing tomography. \citet{Ma08} have shown that a high accuracy
photo-z is needed to avoid any bias on the Dark Energy parameter
estimation. The only way to reach high accuracy photo-z is to plan a
spectroscopic redshift survey to calibrate the photo-z templates, as
in \citet{Ilbert06}.  Ideally, the photo-z calibration survey (PZCS)
would reach the same magnitude and redshift depth as the WL
photometric survey. However, this goal will likely be difficult to
achieve and the CMC emission-line catalog may help to plan the best
spectroscopic strategy. Thus, we design a DEEP-visible-NIR survey (0.6
to 1.7 microns) for weak-lensing, reaching a 3$\sigma$ flux
sensitivity of 5.10$^{-18}ergs\ cm^{-2}s^{-1}$ at 1.2$\mu$m (this can
be achieved with an efficient slitless spectrograph using a 1.5m
telescope, 0.28''/pixel, R$\sim$250 and an exposure time of 240
ksec$\sim$67 hours) as an attempt at calibrating photometric redshifts
for a wide range in redshift and magnitude. The flux sensitivities of
such a survey are represented by the solid red (and medium thickness)
line for a 5$\sigma$ detection and by the solid black (and thin) line
for a 3$\sigma$ detection in {\bf Figure~\ref{jdemsens}}.

{\it (iii)} We also desgin a DEEP-NIR survey having the same
characteritics as the DEEP-visible-NIR survey but covering only the
NIR part (1.0 to 1.7 micron; focussing on the low IR background of
space observation). Our goal here is to evaluate the importance of
having a spectroscopic contribution in the visible wavelength to reach
a high completness.

In {\bf Figure~\ref{jdemsens}}, we also represent the VVDS-DEEP flux
sensitivities in green (and thin) dashed line for a $3\sigma$ detection
and gold (and thick) dashed line for a $5\sigma$ detection. For
reference, the exposure time of the VVDS-DEEP survey is $\sim$10ksec
exposure on a 8m ground-based telescope with a spectroscopic
resolution of R=250.
\subsection{SSR prediction}
\label{subsec:ssr}
In the following sections we will give the basics of photo-z
calibrations survey studies based on the spectroscopic success rate as
a function of redshift and magnitude. A more thorough analysis will be
developed in a forthcoming paper (Jouvel et al 2009 in prep).

\begin{figure}[!ht]
\resizebox{\hsize}{!}{\includegraphics{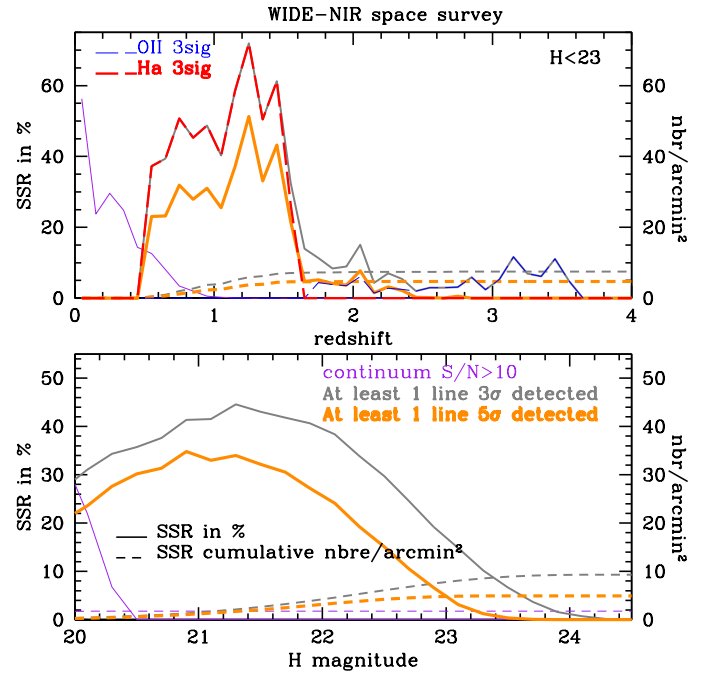}}
\caption{ WIDE NIR space survey SSR.}
\label{BAOssr}
\end{figure}

\subsubsection{SSR for the WIDE survey}
{\bf Figures~\ref{BAOssr}} shows SSR forecasts for a future NIR WIDE
survey. Top panel represents the SSR as a function of redshift and
bottom panel as a function of magnitude. Both panels show the SSR
using a one line at $5\sigma$ detection criterion in orange very
thick, a one line at $3\sigma$ detection criterion in grey thick , and
a continuum S/N$>$10 criterion in purple thin. The solid curves are a
percentage with the total number of objects and the dashed for a
cumulative number of objects by arcmin$^2$. The top panel shows also
the SSR of a 3$\sigma$ detection of the OII emission line in blue
long-dashed thin line and H$\alpha$ in red long-dashed thick line.
The wide survey would aim to cover a large fraction of the sky (10,000
deg$^2$ or more) in order to probe the large scale distribution of
galaxies aiming in particular to measure the baryon acoustic
oscillation with a great accuracy in the redshift range $0.5<z\lesssim
1.5$ (or up to $z\sim 2$ providing the telescope and instrument is
sensitive up to 2$\mu$m). Our SSR prediction, shows that such survey
could easily measure the redshift of more than 4 galaxy/armin$^2$
hence providing the redshift measurement of more than $\sim$100
million galaxies for a survey covering 10,000 deg$^2$. The redshift
identification is essentially based on the H$\alpha$ line detection
thus essentially targeting star-forming galaxies with $H\lesssim
22.5$. The SSR clearly shows that such survey is far from being
complete as the spectroscopic success rate is generally below 40\% for
any magnitude and redshift (for a 5$\sigma$ line detection). Indeed,
there are very few faint galaxies with spectroscopic success,
especially in the redshift ranges which have the strongest need for
photometric redshift calibration : $0<z<0.5$ and
$z>1.5$. Table~\ref{surveys} shows that the SSR also depends strongly
on galaxy type.  Thus, the wide survey {\it can not be used} to
calibrate any photometric redshift catalogue properly.

\subsubsection{SSR for the DEEP survey}

\begin{figure}[!ht]
\resizebox{\hsize}{!}{\includegraphics{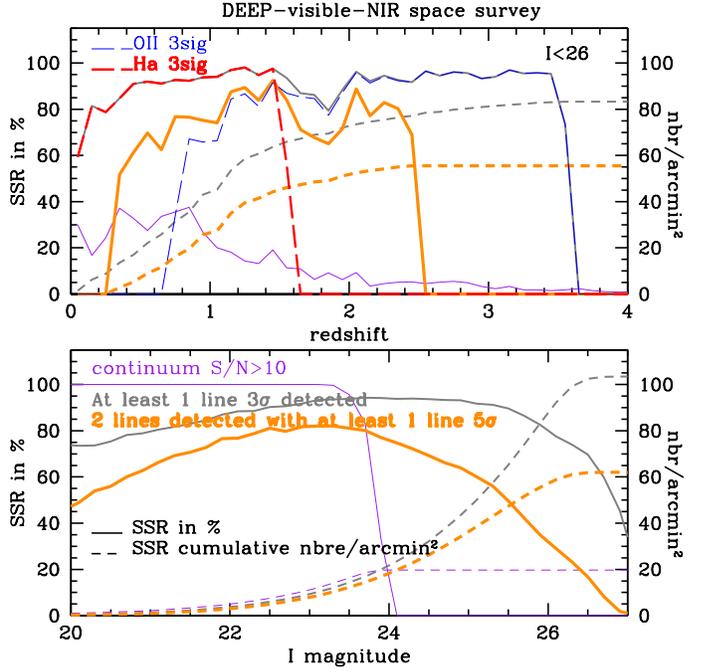}}
\caption{DEEP-visible-NIR space survey SSR.}
\label{jdemssr}
\end{figure}

\begin{figure}[!ht]
\resizebox{\hsize}{!}{\includegraphics{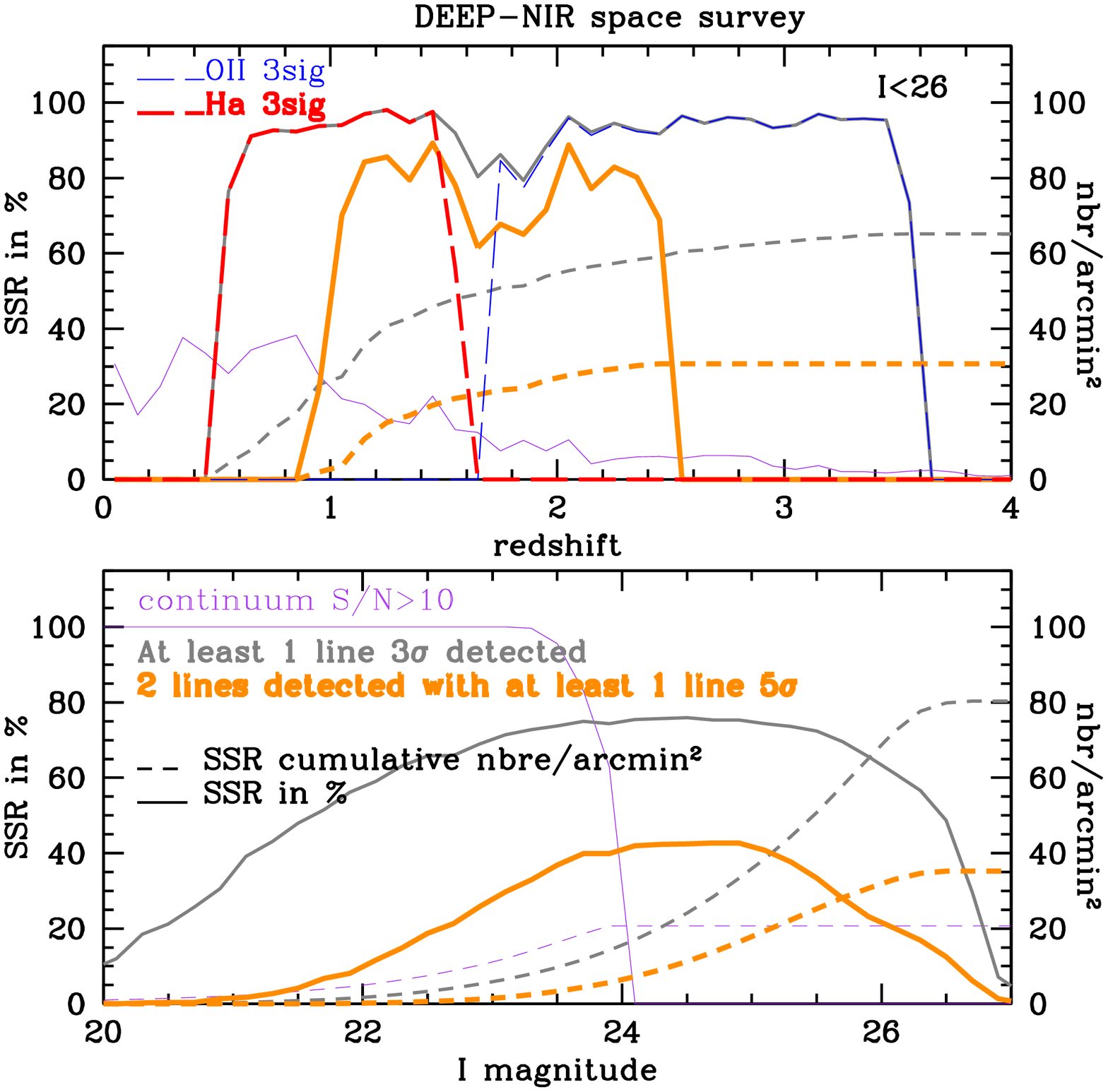}}
\caption{DEEP-NIR only space survey SSR.}
\label{jdemIRssr}
\end{figure}

{\bf Figures~\ref{jdemssr} and ~\ref{jdemIRssr}} show the SSR
prediction for future dark energy surveys which are planing to do
infrared only (DEEP-NIR survey), and visible+infrared spectroscopy
(DEEP-visible-NIR survey) from space. Figure~\ref{jdemsens} shows the
grism flux sensitivities we use in our analysis.  It shows SSR as a
function of redshift [top panel] and magnitude [bottom panel]. Both
panels show the SSR using a 2 lines detection with at least one line
at $5\sigma$ detection criterion in orange very thick, a one line at
$3\sigma$ detection criterion in grey thick, and a continuum S/N$>$10
criterion in purple thin. The solid curves are a percentage with the
total number of objects and the dashed for a cumulative number of
objects by arcmin$^2$. The top panel shows also the SSR of a 3$\sigma$
detection of the OII emission line in blue long-dashed thin line and
H$\alpha$ in red long-dashed thick line.

We assume here that the visible spectroscopy spans 0.6-1$\mu$m see
Figure~\ref{jdemsens}. Because of the limited wavelength coverage, the
$H\alpha$ is the only strong emission line visible for galaxies at low
redshift (z$<$0.5). This explains that the detection criterion based
on 2 lines, does not cover the low redshift range as shown in the top
panel of Figure~\ref{jdemssr}. This is also seen in the bottom panel
of Figure~\ref{jdemssr} where there is a lower SSR for bright
galaxies. However, for these low redshift ($z<0.5$) galaxies, about
30\% have a strong continuum with S/N$>$10. For these bright galaxies,
we should be able to accurately measure a redshift using either the
H$\alpha$ line detection, absorption lines and the shape of the
continuum which will be useful for photometric redshift calibration.

If only the NIR spectroscopy is conducted from space, the survey will
only reach $\sim 35$galaxy per arcmin$^2$ and a maximal SSR of at best
40\% for $I\sim 24$. Using DEEP-NIR spectroscopic survey would help to
calibrate spectra at $0.8<z<2.5$ but would be less efficient below
$z<0.8$. There will be a strong need to conduct visible spectroscopy
to achieve a complete census of the redshift distribution.

Using the DEEP-visible-NIR survey, we should be able to measure a very
secure redshift (2 lines detected at more than 3$\sigma$ with at least
1 emission line at $5\sigma$) for about 60 galaxy per arcmin$^2$ to
$I<26$ and $z\lesssim 2.5$ with a SSR reaching 80\% at best for $I\sim
24$. With this density of sources, a survey of 1 square degree will
provide 220000 spectra. Note, however, that these numbers do not take
into account the crowding of galaxy spectra which is likely a strong
limiting factor above $I\sim24$. There are ways to minimize the impact
of crowding, for example by conducting the slit-less spectroscopic
survey using different orientations, or by using masks to block some
of the light (thus limiting the sky background and reducing the number
of overlapping spectra). These alternatives will be discussed in a
forthcoming paper (Zoubian et al. 2009 in prep. \nocite{Zoubian09}).
Note also that the visible part of the spectroscopy might be conducted
with a dedicated ground-based spectrograph with high multiplexing.
This can probably be especially more efficient for $\lambda<0.75\mu$m
where the sky emission lines are less numerous than in the redder part
of the spectrum.
\begin{table}[!ht]
\caption{Characteristics of the three surveys discussed in the text,
  assuming an efficient 1.5m space telescope with an obscuration of
  0.6m and a total telescope throughput (CCD, mirror, grism) of 70\%
  using a survey efficiency of 75\%. The \%[1line$3\sigma$] Ell
  represent the percentage of Elliptical galaxies spectroscopically
  found using a success criterion of 1 line detected at $3\sigma$
  compared to the total population of Elliptical galaxies. With the
  same criterion, we define the percentage of Sac for early spiral
  galaxies, Sdm for late spiral galaxies, and SB for starburst
  galaxies.}
\begin{tabular}{cccccccc} \hline\hline
& DEEP-vis-NIR & DEEP-NIR & WIDE-NIR \\
probe & WL & WL & BAO \\
$T_{obs}$ & 240ksec & 240ksec & 1200sec\\
$\lambda$ in $\mu$m & 0.6-1.7 & 1-1.7 & 1-1.7 \\
3$\sigma$ at 1.2$\mu$m & $5.10^{-18}$ & $5.10^{-18}$ & $1.10^{-16}$ \\
& & $ergs\ cm^{-2}sec^{-1}$ & \\
area needed & $10deg^2$ & $10deg^{2}$ & $20000deg^{2}$\\
mission time & 0.2yrs & 0.2yrs & 2yrs\\
& & FOV of $0.5deg^{2}$ & \\ 
nbr density (5$\sigma$)& 60 & 35 & 5\\
& & $gal/arcmin^{2}$ & \\
z[1line$3\sigma$] & $0<z<3.5$ & $0.5<z<3.5$ & $0.5<z<1.5$\\
$m_{lim}$ [1line$3\sigma$] & I$\sim$ 27 & I$\sim$ 26.5 & H$\sim$ 22 \\
\%[1line$3\sigma$] Ell & 40 & 27 & 0\\
\%[1line$3\sigma$] Sac & 60 & 51 & 1\\
\%[1line$3\sigma$] Sdm & 77 & 63 & 10\\
\%[1line$3\sigma$] SB & 95 & 74 & 45\\
\hline\hline
\label{surveys}
\end{tabular}
\end{table}
\section{Conclusion}
We have produced simulated deep galaxy catalogs by two different
techniques: one starts with redshift-dependent Schechter luminosity
functions for three galaxy types for $0<z<6$, extrapolating the LFs
derived from GOODS data by \citet{Dahlen05} (the GLFC). The other has
galaxies with redshifts and SEDs taken as the best photo-z fits to
multiwavelength observations of the COSMOS field (the CMC). Both adopt
galaxy size distributions from the COSMOS HST imaging and we assign
emission line strengths using a recipe in agreement with the VVDS-DEEP
data.

Both simulated catalogs do an excellent job of reproducing the $dN/dm$
data ($<20\%$ discrepancies) and color distributions (0.1--0.2 mag
median color discrepancies) of galaxies observed in bands from 0.4 to
2.2 $\mu$m using a comparison with the GOODS and UDF surveys. The CMC
provides an excellent match to the redshift-magnitude and
redshift-color distributions for $I<24$ galaxies in the VVDS
spectroscopic redshift survey.

These simulated catalogs thus pass all our validation tests for use in
forecasting the galaxy ``yields'' of future visible/NIR imaging
surveys, as long as we restrict our analysis to the $I<25.5$--26
galaxies for which the COSMOS HST imaging is highly complete. These
simulated catalogs provide a conservative estimate of the yield for
surveys that go deeper in the visible, or reach $>24$ in the NIR,
because they miss the faint or red galaxies that do not make the $I$-band
cut.

The GLFC has the potential to forecast deeper surveys than the CMC,
but only if we are ready to extrapolate the luminosity functions to
fainter magnitudes, which is probably feasible in the visible bands,
but may be more strongly limited in the near-infrared bands.

In the CMC, the galaxy sizes are directly coming from the HST/ACS
measurement and give us the opportunity to evaluate differences in
terms of possible shape measurements for a ground-based and space
telescopes. A ground based WL survey is limited to use only the
resolved galaxies, and we show that most $I>25$ galaxies are smaller
than the ground-based seeing disk.  Thus the WL depth of ground-based
surveys is basically independent of the survey depth for $I>25$. For a
space survey the number of resolved galaxies is not only dependent on
the PSF but also on the depth of the survey.

Moreover we have also used the CMC and the emission line model to
compare the spectroscopic success rate observed by the VVDS and the
one we can predict based on the CMC line fluxes. In general we obtain
a good agreement between observed success rates and the CMC
predictions.

We have then used the emission line model to explore what will be the
yield in terms of redshift measurements for 2 type of surveys: {\it
  (i)} a WIDE survey motivated by measuring BAO, and {\it (ii)} a deep
survey motivated to calibrate photometric redshifts. We found that the
WIDE survey reaches a density of $4$gal/arcmin$^2$ for galaxies
$H\lesssim 22.5$ in a redshift range of $0.5<z<1.5$ mainly detected
using the $H\alpha$ emission line. Furthermore, in a weak-lensing
perspective, a DEEP survey both visible and NIR is needed to conduct a
photometric redshift calibration of galaxies (PZCS) to be used in a
weak-lensing measurement covering the full magnitude and redshift
range. Indeed such survey reaches a density of 60 gal/arcmin$^2$ for
very secure redshifts down to $I\sim25.5$ with a high completeness of
$\sim$80\%. We note that the visible part of the spectroscopic survey
can be done from the ground, however above $\lambda \sim 0.75 \mu$m a
space survey is likely to be much more efficient than a ground-based
survey (see Figure~\ref{jdemsens}).

We have thus demonstrated here how useful realistic simulated catalogs
are for designing future DE space missions. We will investigate more
precisely details of the WL optimisation survey strategy in a
forthcoming paper (Jouvel et al. 2009, in prep. \nocite{Jouvel09}).

\begin{acknowledgements}
We acknowledge useful discussions with members of the COSMOS and SNAP
collaborations. Stephanie Jouvel thanks CNES and CNRS for her PhD
studentship. Jean-Paul Kneib thanks CNRS for support. Gary Bernstein
is supported by grant AST-0607667 from the National Science
Foundation, Department of Energy grant DOE-DE-FG02-95ER40893, and NASA
grant BEFS-04-0014-0018.
\end{acknowledgements}

\bibliographystyle{aa} \bibliography{biblio}

\end{document}